# WEAKLY-BOUND THREE-BODY SYSTEMS WITH NO BOUND SUBSYSTEMS

Jérôme Goy[(1)], Jean-Marc Richard[(1),(2)], and Sonia Fleck[(3)],
[(1)]*Institut des Sciences Nucléaires*
*53, avenue des Martyrs, 38026 Grenoble, France*
[(2)]*Institut für Theoretische Kernphysik*
*Rheinische Friedrich-Wilhelms Universität*
*Nußallee 14-16, D-53115 Bonn, Germany*
[(3)]*Institut de Physique Nucléaire*
*43, boulevard du 11 Novembre 1918, 69622 Villeurbanne, France*
(August 15, 1995)

## Abstract

We investigate the domain of coupling constants which achieve binding for a 3-body system, while none of the 2-body subsystems is bound. We derive some general properties of the shape of the domain, and rigorous upper bounds on its size, using a Hall–Post decomposition of the Hamiltonian. Numerical illustrations are provided in the case of a Yukawa potential, using a simple variational method.

03.65.Ge, 21.45.+v, 21.80.+1, 21.60.Gx

Typeset using REVTEX

# I. INTRODUCTION

In 3-dimensional quantum mechanics, attractive short-range potentials do not always produce bound states. For given constituent masses, a minimal strength is required. Equivalently, particles should be heavy enough to experience binding in a given potential.

More interesting perhaps is the observation that 3-body systems can be bound by pairwise potentials which are not attractive enough to bind the corresponding 2-body subsystems. In this paper, we follow Ref. [1] and call such a 3-body system a "Borromean state", after the Borromean rings, which are interlaced in such a subtle topological way, that if any of them is removed, the other two would be unlocked.

Some Borromean systems have been known for several years in nuclear physics. For instance, $^6$He is stable against spontaneous dissociation, while $^5$He is unstable. If one neglects the internal structure of the $\alpha$ (= $^4$He) particle, a fairly honest approximation, $^6$He is a bound $(\alpha - n - n)$ system, while neither $(\alpha - n)$ nor $(n - n)$ is bound. Thus $^6$He is Borromean. In atomic physics, Borromean or nearly-Borromean states are also expected, as the potential between neutral atoms has attractive parts, but remains weak, so that many 2-atom systems are unbound or very weakly bound. For instance, Efimov [3] pointed out that some models for the $^4$He–$^4$He interaction have an unbound ($^4$He)$_2$ dimer and a bound ($^4$He)$_3$ trimer.

There are also $N$-body Borromean states with $N > 3$, whose all $N'$-body subsystems, $N' < N$, are unbound. One finds even more complicated situations, with for instance $(N - 1)$-body subsystems unstable, and some $(N - 2)$-body ones stable. An example is $^8$He considered as a $(\alpha - n - n - n - n)$ system, as $^6$He is stable and $^7$He is not.

There are at least two well-known and extensively-studied quantum phenomena supporting the idea that 3-body systems might be more easily bound than the 2-body ones: the Efimov effect and the Thomas collapse.

The Efimov effect [4] occurs for a coupling constant close to that giving a zero-energy 2-body bound state, or, say, an infinite scattering length. Many loosely bound 3-body states exist in this limit. Recent papers have proposed new derivations of the Efimov effect, or new points of view for its understanding [5].

The Thomas collapse [6] is the observation that when the range of the potential decreases, the ratio $E_3/E_2$ of 3-body to 2-body ground-state energies becomes very large. This means there is much more binding per particle in the 3-body than in the 2-body system. It is remarkable that Thomas was able to set a limit on the the range of nuclear forces, as early as in 1935, by comparing the energies of 3-nucleon and 2-nucleon bound states.

Our aim is to investigate the domain of coupling constants which produce Borromean systems. If one considers for instance a Yukawa potential, the critical coupling $g_3$ to bind three identical particles is by around 20% smaller than the coupling $g_2$ necessary to bind two particles. The question is whether $g_3/g_2$ can be made very small by suitable tuning of the shape of the potential. We shall see that $g_3/g_2$ cannot be made smaller than 2/3. In less symmetric situations where we are dealing with two or three coupling constants, we shall find upper bounds on the domain for Borromean binding. Our study is somewhat complementary to papers on the Efimov effect, showing the richness of the 3-body spectrum near $g = g_2$ [4,5], or looking at the ground-state energy for $g = g_2$ [7]: we wish to determine how far one can bind below $g_2$.

This paper is organized as follows. Sec. II contains brief reminders on the 2-body case.

In Sec. III, we derive bounds on the size of the domain of coupling constants for Borromean binding, and some convexity properties of this domain. In Sec. IV are presented numerical results obtained by a variational calculation applied to Yukawa potentials. A few generalizations are proposed in Sec. V, as well as some related problems. Some of the results were already presented in Ref. [8], and are repeated here with more details. A brief summary is provided in Sec. VI.

## II. BINDING 2-BODY SYSTEMS

Let us consider two particles of masses $m_1$ and $m_2$ interacting through a translation invariant potential $V$, which will be usually of the type $V(r)$, where $r = |\vec{r}| = |\vec{r}_1 - \vec{r}_2|$, though the local character is not always required. We are interested in the translation-invariant part of the Hamiltonian,

$$\tilde{H}_2(\mu, g) = \frac{\vec{p}^2}{\mu} + gV, \qquad (2.1)$$

where the relative momentum $\vec{p}$ is conjugate of $\vec{r}$, and $\mu$ twice the reduced mass. Even if the potential has attractive parts ($V < 0$), binding is not always achieved. A minimal value is often required for the strength $g$

$$\mu g \geq g_2 . \qquad (2.2)$$

Exceptions include potentials with a slow asymptotic decrease, like the Coulomb potential [9]. From now on, we shall restrict ourselves to potentials which vanish very rapidly at large separation, so that Eq. (2.2) holds with a finite value of $g_2$. Consider for instance, a Yukawa potential: one can fix the range and the reduced mass by rescaling, and restrict oneself to $V = -\exp(-r)/r$ in the case where $\mu = 1$. A simple argument by Dyson and Lenard [10] shows that $g_2 > \sqrt{2}$. The value of $g_2$ is available in the literature [11,12], $g_2 \simeq 1.6798$.

Within variational methods, it is not exactly the same art to compute the ground state $E_2$ with high accuracy in a regime $g > g_2$ where stability is ensured, and to compute very precisely the coupling $g_2$ where $E_2 = 0$; see, for instance, Ref. [12]. The difficulty comes from the slow onset of the amount of binding near $g_2$ [13], of the type

$$E_2(g) \propto -(g - g_2)^2 . \qquad (2.3)$$

To estimate $g_2$, one can also integrate numerically the radial equation for S-states at energy $E_2 = 0$, using for instance the algorithm proposed by Hartree [14], and look at which coupling $g = g_2$ the radial wave function starts exhibiting a node at large distances. Great accuracy can be obtained, as one can check with potentials for which exact results are known [15].

## III. GENERAL PROPERTIES OF THE DOMAIN OF STABILITY

### A. Symmetric case

Let us consider the simplest case of three identical bosons with a mass set to $m = 1$ to fix the scale. As a first example, we take a Yukawa potential $-g \sum_{i<j} \exp(-r_{ij})/r_{ij}$, with $r_{ij} = |\vec{r}_i - \vec{r}_j|$. With standard numerical methods, on which more later, one can calculate the critical coupling $g_3$ for three-body binding. One finds $g_3/g_2 \simeq 0.804$, i.e., a 20% window for Borromean binding. If one repeats the computation of this ratio of critical couplings with other simple potentials, one gets comparable results, for instance $g_3/g_2 \simeq 0.801$ for an exponential $V = -\exp(-r)$, 0.794 for a Gaussian $V = -\exp(-r^2)$, and 0.806 for a Hulthen potential $V = -1/(\exp(r) - 1)$ [16]. Such quasi-universality was already noticed in [17], where the 3-body ground-state energy was plotted against the 2-body scattering length. It is not too surprising, as in the weak-binding limit we are dealing with, most of the wave function lies outside the potential well, which is not very much probed.

One might however address the question whether changing drastically the shape of $V(r)$ could result into much smaller or much larger values of $g_3/g_2$.

Potentials with a large repulsive core, such as the Morse interaction with a minimum located at large distance, could lead to $g_3/g_2$ larger than 0.8. This is currently under study and will be reported elsewhere [18].

On the other hand, one can show that $g_3/g_2$ is bounded below, namely

$$\frac{g_3}{g_2} \geq \frac{2}{3} \,. \tag{3.1}$$

This inequality is nearly saturated for a modified harmonic oscillator, which is $V(r) = V_0 + r^2$, $V_0 < 0$ at short distances, and cut off at very large $r$ so that $V \to 0$.

The proof consists of a simple modification of the Hall–Post inequalities [19] which relate 3-body to 2-body energies at fixed coupling, providing a simple lower bound for the former. In the course of studies on the stability of matter or on the quark model, a weaker version has been proposed [10,20,21], where the energy of the centre-of-mass motion was not properly removed. The optimal form of this inequality, with saturation in the case of harmonic forces, was recently rediscovered [22], applied to a comparison of meson and baryon masses in the quark model [23], and generalized to the case of unequal constituent masses [24].

From the 3-boson Hamiltonian

$$H_3 = \sum_{i=1}^{3} \frac{\vec{p}_i^2}{2m} + g \sum_{i<j} V(r_{ij})$$
$$= \frac{(\sum \vec{p}_i)^2}{6m} + \widetilde{H}_3 \,, \tag{3.2}$$

one extracts a translation-invariant part $\widetilde{H}_3$ which can be written as

$$\widetilde{H}_3 = \sum_{i<j} \frac{2}{3m} \left(\frac{\vec{p}_i - \vec{p}_j}{2}\right)^2 + gV(r_{ij}), \tag{3.3}$$

to exhibit momenta which are canonical conjugate to the relative distances $\vec{r}_i - \vec{r}_j$. In short,

$$\widetilde{H}_3(m,g) = \sum_{i<j} \widetilde{H}_2^{(ij)}(3m/2, g) = \frac{2}{3} \sum_{i<j} \widetilde{H}_2^{(ij)}(m, 3g/2). \tag{3.4}$$

If one saturates the above operator identity with the ground eigenstate of $\widetilde{H}_3$, and apply the variational principle to the matrix elements of $\widetilde{H}_2$, in a regime of given coupling $g$ (large enough, so that every Hamiltonian binds), one gets

$$E_3(m,g) \geq 3E_2(3m/2, g) = 2E_2(m, 3g/2), \tag{3.5}$$

which is the simplest form of the Hall–Post inequality [19,22].

By a similar reasoning, one would never get a 3-body bound state, corresponding to $\langle \widetilde{H}_3 \rangle < 0$ with the appropriate wave function, as long as the sub-Hamiltonians do not bind, i.e., $\langle \widetilde{H}_2 \rangle > 0$ for any state. Thus $3g/2 \geq g_2$ is required, q.e.d.

### B. Partially symmetric case

We now assume that only particles 1 and 2 are identical. The Hamiltonian is of the form

$$H_3 = \frac{\vec{p}_1^2}{2} + \frac{\vec{p}_2^2}{2} + \frac{\vec{p}_3^2}{2M} + g\frac{1+M}{2M}(V_{23} + V_{31}) + g'v_{12}, \tag{3.6}$$

where the potentials $v_{12}$ and $V_{ij}$ are attractive, and of short range. The potential functions $V$ and $v$ need not being identical. The normalization is chosen such that $m_1 = m_2 = 1$, $g' = 1$ for the critical binding to bind (1,2), and also $g = 1$, the critical coupling to bind masses $(1, M)$ with potential $V$. The domain for Borromean binding is thus to be found inside the unit square ($g \leq 1, g' \leq 1$), as schematically pictured in Fig. 1.

The inner part is the region of no binding, where the minimum of the Hamiltonian is $\min[H_3] = 0$, i.e., the beginning of the continuum. Since the coupling constants enter the Hamiltonian linearly, if one considers two points $P(g, g')$ and $\widetilde{P}(\tilde{g}, \tilde{g}')$ in the no-binding region, and an intermediate point $Q = \lambda P + (1 - \lambda)\widetilde{P}$, with $0 \leq \lambda \leq 1$, then

$$H_3(Q) = \lambda H_3(P) + (1 - \lambda)H_3(\widetilde{P}), \tag{3.7}$$

and [25]

$$\min[H_3(Q)] \geq \lambda \min[H_3(P)] + (1 - \lambda)\min\left[H_3(\widetilde{P})\right] = 0. \tag{3.8}$$

This means the instability region is a convex domain.

The region of binding cannot extend up to the $g = 0$ axis, as a bound (1,2) pair needs a minimum of attraction to remain linked to the third particle.

The behaviour of the frontier of stability near the $g' = 0$ axis, where particles 1 and 2 only interact with 3, depends on the value of $M$. In the case of an infinitely heavy nucleus $M = \infty$, one strictly needs $g \geq 1$, with our normalization. Hence the frontier ends at the lower corner ($g = 1, g' = 0$) of the unit square. For a finite-mass nucleus ($M < \infty$), the frontier might end at some point $(g_0, 0)$ with $g_0 < 1$. An heuristic argument is the following. If particle 1 is assumed to be bound to 3, then particle 2 interacts with a kernel of mass $(1 + M)$, to which binding is easier than with $M$ alone. Now 2 linked to 3 is heavier than 3 alone, and this might justify the hypothesis that 1 is bound. Whether or not $g_0 < 1$ is related to the discussion on the sign of the correction to the ground-state energy of Helium due to the motion of the nucleus. Using explicitly $\vec{p}_3 = -\vec{p}_1 - \vec{p}_2$, the $g' = 0$ case reads [25,26]

$$H_3 = \left(\frac{1}{2} + \frac{1}{2M}\right) \sum_{i=1,2} \left[\vec{p}_i^2 + gV_{i,3}\right] + \frac{1}{M}\vec{p}_1 \cdot \vec{p}_2 . \tag{3.9}$$

Without the Hughes–Eckart term, one gets a simple factorizable solution, and stability requires $g \geq 1$. The Hughes–Eckart term is repulsive for the ground state of the He atom, due to the anticorrelation between the electrons [25,26]. Here, the analog of the electrostatic repulsion between the electrons vanishes ($g' = 0$), and the Hughes–Eckart term tends to become attractive, to provide some stability for $g < 1$.

We now derive an upper bound on the size of the domain of stability. The strategy is the same as in the symmetric case: we split the Hamiltonian into simple pieces, and look at whether the sub-Hamiltonians can reach negative values. To show that too simple a decomposition of the Hamiltonian is not sufficient to provide a satisfactory bound, let us consider two examples. The first one is the case $M = \infty$ [8]. A simple decomposition is

$$2H_3 = \left[\alpha \vec{p}_1^2 + gV(r_1)\right] + \left[\alpha \vec{p}_2^2 + gV(r_2)\right] + \left[(1-\alpha)\left(\vec{p}_1^2 + \vec{p}_2^2\right) + g'v(r_{12})\right], \qquad (3.10)$$

for any $0 \leq \alpha \leq 1$. To get $\langle H_3 \rangle < 0$, one needs at least one of the square brackets having a negative expectation value. This excludes the triangle $\{g \leq \alpha, g' \leq (1-\alpha)\}$, shown in Fig. 2. However, we are touching the $g = 0$ axis, and do not get a strictly convex domain.

Another example is the equal-mass case $M = m$. One can still use the decomposition (3.3) of the symmetric case, rewritten as

$$H_3 = \frac{(\vec{p}_1 + \vec{p}_2 + \vec{p}_3)^2}{6} + \sum_{i=1,2} \left[\frac{2}{3}\left(\frac{\vec{p}_i - \vec{p}_3}{2}\right)^2 + gV_{i3}\right] + \left[\frac{2}{3}\left(\frac{\vec{p}_1 - \vec{p}_2}{2}\right)^2 + g'v_{12}\right]. \qquad (3.11)$$

Clearly, $H_3$ would never reach negative expectation values as long as each bracket remain positive. The inner square domain

$$g \leq \frac{2}{3}, \qquad g' \leq \frac{2}{3} \qquad (3.12)$$

is thus excluded (see Fig. 3). Again, this bound is rather crude, since it does not exclude some points on the $g = 0$ axis, and the actual frontier is not expected to be a flat function of $g$ or $g'$.

We thus have to look at a more general decomposition of the Hamiltonian. For arbitrary mass $M$, we rewrite (3.6) in the form [24]

$$H_3 = (\vec{p}_1 + \vec{p}_2 + \vec{p}_3) \cdot (b\vec{p}_1 + b\vec{p}_2 + b'\vec{p}_3)$$
$$+ \sum_{i=1,2} \left[a\left(\frac{\vec{p}_i - x\vec{p}_3}{1+x}\right)^2 + g\frac{1+M}{2M}V_{i3}\right] + \left[a'\left(\frac{\vec{p}_1 - \vec{p}_2}{2}\right)^2 + g'v_{12}\right]. \qquad (3.13)$$

The momenta in the sub-Hamiltonians are normalized to be conjugate to the relative distances $\vec{r}_i - \vec{r}_j$. Note that the first term might differ from the kinetic energy of the c.o.m. motion, but still vanishes for the ground state of $H_3$. For any given $x \geq 0$, one can identify (3.13) with the original Hamiltonian (3.6), and calculate $a$ and $a'$ as functions of $x$ (one can also calculate $b$ and $b'$, but their values are irrelevant). One gets

$$a = \left(\frac{1+x}{1+2x}\right)^2 \left(\frac{1}{2} + \frac{1}{M}\right)$$
$$a' = \frac{4}{(1+2x)^2}\left(x(1+x) - \frac{1}{2M}\right). \qquad (3.14)$$

From (3.13), $H_3$ will not bind as long as $g(1+M)/(2M) \le a$ and $g' \le a'$. This gives an improved lower limit for the frontier of stability, which one can write as

$$g = \frac{M+2}{4(M+1)}(1+t)^2$$
$$g' = 1 - \frac{M+2}{M}t^2 \,. \tag{3.15}$$

after a suitable change of variable. This is an arc of parabola. The domain of variation is first $0 \le x < \infty$ when writing (3.13), corresponding to $0 \le t \le 1$, but it is further restricted by the requirement that the inverse masses $a$ and $a'$ should be positive.

The $M = \infty$ case is shown in Fig. 2. It excludes the low $g$ part ($g < 1/4$), and exhibits the appropriate convexity. The lower bound for $M = 1$ is displayed in Fig. 3. It ends at $g' = 0$ and $g = 1/2 + \sqrt{3}/4 \simeq 0.933$, leaving at most a 7% margin for Borromean binding when the interaction $v_{12}$ is switched off.

One notices an horizontal behavior when reaching the $g' = 1$ threshold, and the slope $dg'/dg = -2$ on the symmetry axis $g = g'$, in the case where $M = 1$. The same slope should be observed for the actual frontier in the case of equal masses ($M = 1$) and identical potentials ($V = v$), since one can split the Hamiltonian into

$$H = \sum_i \frac{\vec{p}_i^2}{2} + \frac{2g+g'}{3}(V_{12} + V_{23} + V_{31}) + \frac{g-g'}{3}(V_{23} + V_{31} - 2V_{12}), \tag{3.16}$$

so that the last term, of mixed permutation symmetry, contributes at second order only when treated in perturbation about the symmetric term. Hence the energy is mostly a function of $(2g + g')$, and the frontier has a slope $-2$ near $g = g'$.

### C. General case

The generalization to three different couplings $h$, $g$, and $k$ is rather straightforward, so we shall be rather brief here, and mostly give the results. The Hamiltonian is

$$H_3 = \frac{\vec{p}_1^2}{2m_1} + \frac{\vec{p}_2^2}{2m_2} + \frac{\vec{p}_3^2}{2m_3} + \frac{g}{m_{23}}u_{23} + \frac{h}{m_{31}}v_{31} + \frac{k}{m_{12}}w_{12}, \tag{3.17}$$

where $m_{ij} = 2m_i m_j/(m_i + m_j)$ is twice the reduced mass, so that $g = 1$ is the critical coupling for binding $m_2$ and $m_3$ in potential $u$, and similarly for $h$ and $k$.

The domain of Borromean binding is thus restricted inside the unit cube ($g \le 1, h \le 1, k \le 1$). Its complement, the domain of no-binding, is convex.

The frontier has of course some symmetries if the masses are equal, and the potential terms $u$, $v$ and $w$ have identical functional dependence. For instance, its normal vector is parallel to (1,1,1) at the point where $g = h = k$.

The Hamiltonian (3.17) can be split into [24]

$$H_3 = (\vec{p}_1 + \vec{p}_2 + \vec{p}_3) \cdot (b_1\vec{p}_1 + b_2\vec{p}_2 + b_3\vec{p}_3)$$
$$+ \left[ a_3 \left( \frac{\vec{p}_1 - x_3\vec{p}_2}{1 + x_3} \right)^2 + \frac{k}{m_{12}}v_{12} \right] + \ldots, \tag{3.18}$$

where the inverse masses are given by

$$a_3 = \frac{(1+x_3)^2}{2} \frac{x_1(1+x_2)/m_2 + (1+x_1)/m_1 - x_2/m_3}{(1+x_1+x_3x_1)(1+x_3+x_2x_3)}, \tag{3.19}$$

and circular permutations. The inverse mass $a_3$ is of course invariant under simultaneous $m_1 \leftrightarrow m_2$, $x_1 \leftrightarrow x_2^{-1}$ and $x_3 \leftrightarrow x_3^{-1}$ exchanges. From (3.19), the ground state of $H_3$ fulfills

$$E_3(H_3) \geq E_2\left(a_1, \frac{g}{m_{23}} u_{23}\right) + \cdots, \tag{3.20}$$

with obvious notations. Any optimization of the r.h.s. of the above inequality leads to equations of the type

$$\sum_{i=1}^{3} \frac{dE_2(a_i)}{da_i} \frac{\partial a_i}{\partial x_j} = 0, \tag{3.21}$$

which cannot be satisfactorily fulfilled unless the Jacobian determinant $\det(\partial a_i / \partial x_j)$ vanishes, i.e.,

$$x_1 x_2 x_3 = 1, \tag{3.22}$$

as shown in [24]. We shall always keep this condition satisfied.

The frontier of stability is bounded below by the surface

$$g = m_{23} a_1(x_1, x_2, x_3), \quad h = m_{31} a_2, \quad k = m_{12} a_3. \tag{3.23}$$

This limiting surface is shown in Fig. 4, in the case of equal masses $m_i = 1$. This particular choice provides the surface with symmetries. The largest departure from the cube occurs on the symmetry axis, where $g = h = k$. The wave function can be written in three possible ways as a product of a 2-body cluster times a third particle, schematically $\Psi = \sum \alpha_k [(i,j), k]$. On the symmetry axis, all $[(i,j), k]$ components are degenerate in energy. They can thus experience sizeable constructive interferences, which lower the energy of $\Psi$.

## IV. NUMERICAL RESULTS

To illustrate how likely is the occurrence of Borromean binding, and how realistic are the lower bounds derived in the previous section, we have explicitly computed the stability frontier for a Yukawa potential.

To cross-check the computation, several methods of solving the 3-body problem, have been used. As the comparison of the numerical results is a little tedious, it will be restricted to the symmetric case.

### A. Methods

We consider here three identical bosons, with mass $m_i = 1$, interacting through a pairwise local potential $g \sum_{i<j} v(r_{ij})$. In the case of a Yukawa interaction, $v(r) = -\exp(-r)/r$, some numerical results obtained with different methods are shown in Table I. These are some binding energies for given coupling $g > g_3$, and estimates of the critical coupling $g_3$.

The methods used in these Tables are:

1. The hyperscalar approximation. The wave function is restricted to be of the type $\Psi = \psi(R)/R^{5/2}$, where $R^2 = 2(\sum_{i<j} r_{ij}^2)/3$, resulting into the 2-body-like radial equation

$$\psi''(R) - \frac{15}{4R^2}\psi(R) + [E - gV_0(R)]\psi(R) = 0, \tag{4.1}$$

where

$$V_0(R) = \frac{48}{\pi} \int_0^{\pi/2} \sin^2\varphi \cos^2\varphi \, v(R\sin\varphi) \mathrm{d}\varphi, \tag{4.2}$$

This is the first step of a more systematic expansion into generalized partial waves [27]. The method can be adapted to the case of unequal masses.

2. The Feshbach–Rubinow method [28] is rather similar, except that $R = (\sum_{i<j} r_{ij})/2$, and $E \to 15E/14$ in the radial equation (4.1). The projection of the potential now reads

$$V_0(R) = \frac{24}{R^5} \int_0^R \left[ R^2 x^2 - Rx^3 + x^4/6 \right] v(x) \mathrm{d}x, \tag{4.3}$$

It can be generalized to unequal masses; see, for instance [7].

3. A Gaussian expansion. If $\vec{\rho}$ and $\vec{\lambda}$ are two Jacobi coordinates describing the relative motion (the specific choice does not matter), the wave function is searched as

$$\Psi = \sum_{i=1}^{G} k_i \left[ \exp -(a_{11}\vec{\rho}^{\,2} + a_{22}\vec{\lambda}^{\,2} + 2a_{12}\vec{\rho}\cdot\vec{\lambda}) + \cdots \right], \tag{4.4}$$

where the dots mean (in general) 5 terms deduced by permutation, to make each bracket explicitly symmetric. The parameters are optimized numerically. This method can be generalized to unequal masses, more than 3 particles, etc., and is widely used in quantum chemistry [29], nuclear physics [30], etc.

4. An expansion in terms of exponentials on the distances, namely

$$\Psi = \sum_{i=1}^{G} k_i \left[ \exp -(a_i r_{23} + b_i r_{31} + c_i r_{12}) + \cdots \right], \tag{4.5}$$

with a similar symmetrization within each bracket. This method is more accurate, as it better accounts for the long-range behavior. It was used in several pioneering paper on few-body systems in atomic physics [31]. It also works for unequal masses. However, with more than $N = 3$ particles, its use becomes rather difficult. Detailed investigations of the convergence properties for those 3-body calculations will be reported elsewhere [32].

5. The Faddeev method in coordinate space [27], restricted to the lowest ($l = 0$) angular momentum for the pairs in the sub-amplitudes.

Comments are in order.

i) The Feshbach–Rubinow method works better than the hyperscalar approximation, for these short-range potentials. A similar conclusion was reached in [33]. This contrasts with the case of the confining potentials one uses in simple quark models of baryon. For instance, with a pairwise linear potential $v(r) = r$ replacing our Yukawa, one would obtain a ground-state energy $E = 6.2089$ with Feshbach–Rubinow, which is worse than the $E = 6.1348$ [27] of the hyperscalar approximation[1].

*ii*) The Gaussian expansion requires many terms ($G$ large) to become efficient. with just a few terms, as in the $G = 4$ case shown here, it is largely superseded by the exponential type of expansion.

*iii*) The Faddeev method is rather efficient in this weak-binding regime, as it incorporates a minimal amount of 2-body correlations in its wave function.

### B. Results

The numerical calculations have been carried out using the variational expansion into exponentials, with $G = 4$ terms, and some explicit symmetrization in particular cases where some masses and couplings are equal.

The results presented below correspond to rescaled distances and couplings. Namely, we rewrite the Hamiltonian

$$H = \sum_i \frac{\vec{P}_i^2}{2M_i} - \sum_{i<j} G_{ij} \frac{e^{-R_{ij}/R_o}}{R_{ij}} = \frac{\hbar^2}{M_1 R_o^2} \left[ \sum_i \frac{\vec{p}_i^2}{2m_i} - \sum_{i<j} \frac{g_{ij}}{m_{ij}} (g_2 \frac{e^{-r_{ij}}}{r_{ij}}) \right] \quad (4.6)$$

so that, within the bracket, $\hbar = m_1 = 1$, and $g_{ij} = 1$ is the critical coupling to bind a $(i,j)$ pair of reduced mass $m_{ij}/2$. The correspondence is $g_{ij} = M_1 R_o m_{ij}/(\hbar^2 g_2)$, with $g_2 = 1.6798$.

In the case where particles 1 and 2 are identical, corresponding to the Hamiltonian (3.6), we obtain the frontier shown in Fig. 5 for and infinite third mass $M$, and Fig. 6 for $M = 1$, where a comparison is done with the rigorous limit. Typically, the domain of actual Borromean states covers around 2/3 of the area allowed by the rigorous bounds. Other investigations have shown that the curves corresponding to exponential, Gaussian or similar potentials are almost identical to those ones [32].

Fig. 7 corresponds to three equal-mass particles, but with different couplings between them. It has to be compared with the lower bound shown in Fig. 4. As we have not pushed very far the variational computation, this surface should be considered somewhat as an upper bound, the actual frontier lying perhaps a little inside.

The window for Borromean binding can be measured by the distance from this frontier surface to the unit cube. The largest window occurs in the symmetric case $g = h = k$. In this case, the three possible decompositions of the wave function into a 2-body cluster and a particle are degenerate, and thus interfere maximally.

---

[1]The hyperscalar approximation is exact if the total potential energy is a function of $\sum r_{ij}^2$, but the Feshbach–Rubinow method remains an approximation for a total potential $V = \sum r_{ij}$ or a function of $\sum r_{ij}$.

# V. GENERALIZATIONS AND RELATED PROBLEMS

Most of the numerical illustrations, in the previous section, deal with equal masses. The influence of the masses deserves further investigations. Consider again the case where two particles are identical. A comparison of the rigorous bounds on $(m, m, M)$ stability is done in Fig. 8, for the $M \ll m$, $M = m$, and $M \gg m$ cases. In the upper-right corner, the window for Borromean binding seems larger in the $M \gg m$ case. This corroborates the hierarchy found in Ref. [7], where the binding energy at $g = g' = 1$ is computed for several values of the ratio $M/m$. The influence of the ratio $M/m$ on the 3-body spectrum was also studied by Efimov [34] and Fonseca and Shanley [35].

For most of the investigations presented in the previous sections, in particular the variational calculations, the cases with more than $N = 3$ particles would require more efforts. However, some of the rigorous bounds can be easily generalized. Some examples are given below.

Consider first the case of $N$ identical bosons, interacting with a short-range potential. A straightforward generalization of (3.4) to $N$ bosons is [22]

$$\widetilde{H}_N(m, g) = \frac{1}{N-2} \sum_{k=1}^{N} \widetilde{H}_{N-1}^{[k]}\left(\frac{Nm}{N-1}, g\right) \tag{5.1}$$

where the superscript in $\widetilde{H}_{N-1}^{[k]}$ means that the $k$-th particle is omitted. Saturating with the ground state of $\widetilde{H}_N$ shows that stability requires [8]

$$g_N \geq \frac{N-1}{N} g_{N-1}, \tag{5.2}$$

i.e. $Ng_N$ increases with $N$. For $N = 4$, a crude variational calculation [8] gives $g_4/g_2 \leq 0.67$ for a Yukawa potential, i.e., once $g_3/g_2$ is subtracted, at least a 13% window for a genuine 4-body Borromean state.

Another interesting situation deals with $p$ identical bosons, of mass set to $m = 1$, in the field of a static source. For $p = 2$, it corresponds to the $M = \infty$ in subsection (III B), from which we take the same notation, $g$ for the particle–source coupling, and $g'$ for the interparticle coupling, both normalized to $g$ or $g' = 1$ for 2-body binding. The following results can be established [8]. First, all stability curves $g' = g'_p(g)$ end at the same point $(g = 1, g' = 0)$ corresponding to independent binding of each particle around the source. Secondly, the Hamiltonians

$$H_p = \sum_{i=1}^{p} \left[\frac{\vec{p}_i^2}{2} + gV_i\right] + \sum_{i<j} v_{ij} \tag{5.3}$$

fulfill the identity

$$H_p(g, g') = \sum_{j=1}^{p} H_{p-1}^{(j)}(g, \frac{p-1}{p-2}g'), \tag{5.4}$$

where $H_{p-1}^{(j)}$ denotes the $(p-1)$-body Hamiltonian with the $j$-th particle being removed. By a now familiar reasoning, binding of $H_p$ requires that the $H_{p-1}^{(j)}$ have their couplings in the stability region, i.e.,

$$g'_p(g) > g'_{p-1}\left(\frac{p-1}{p-2}g\right). \tag{5.5}$$

Fig. 9 is a new drawing of Fig. 2, with the lower limit $g'_2$ for binding $p=2$ particle around a static source, and the lower bound one deduces by the affinity (5.5) for the curve $g'_3$ with $p=3$ bosons. It is presumably rather crude, especially near the upper left corner where $g=0$ and $g'=1$. The decomposition (5.4) hardly accounts for the possibility of forming a 3-boson bound state, of mass $m'=3$, for some $g'<1$, and then binding this mass $m'$ to the source.

Simple results can also be found for another 4-body case, with 2 identical bosons of mass $m$ interacting with two other identical bosons of mass $M$ [8][2]. Let us denote the couplings $g_{mm}$, $g_{MM}$ and $g_{mM}$, all normalized to $g=1$ for 2-body binding. The Hamiltonian

$$H_4 = \sum_{i=1,2}\frac{\vec{p}_i^2}{2m} + \sum_{j=3,4}\frac{\vec{p}_j^2}{2M} + \frac{g_{mm}}{m}u_{12} + \frac{g_{MM}}{M}v_{34} + g_{mM}\frac{m+M}{2mM}\sum_{i,j}w_{ij}, \tag{5.6}$$

can be rewritten as

$$\begin{aligned}H_4 &= (\vec{p}_1+\vec{p}_2+\vec{p}_3+\vec{p}_4)\cdot(b\vec{p}_1+b\vec{p}_2+b'\vec{p}_3+b'\vec{p}_4)\\ &+a_{12}\left(\frac{\vec{p}_1-\vec{p}_2}{2}\right)^2 + \frac{g_{mm}}{m}u_{12} + a_{34}\left(\frac{\vec{p}_3-\vec{p}_4}{2}\right)^2 + \frac{g_{MM}}{M}v_{34}\\ &+\sum_{i,j}\bar{a}[\alpha\vec{p}_i - (1-\alpha)\vec{p}_j]^2 + g_{mM}\frac{m+M}{2mM}w_{ij}\end{aligned} \tag{5.7}$$

with the inverse masses given by

$$\begin{aligned}a_{12} &= \frac{1}{m}(1-\alpha^2) - \frac{1}{M}\alpha^2,\\ a_{34} &= -\frac{1}{m}(1-\alpha)^2 + \frac{1}{M}\alpha(2-\alpha),\\ \bar{a} &= \frac{1}{4m} + \frac{1}{4M}.\end{aligned} \tag{5.8}$$

This shows that $H_4$ cannot support a bound state if simultaneously

$$\begin{aligned}g_{mm} &\leq 1-\alpha^2 - (m/M)\alpha^2,\\ g_{MM} &\leq -(M/m)(1-\alpha)^2 + \alpha(2-\alpha),\\ g_{mM} &\leq 1/2.\end{aligned} \tag{5.9}$$

Interestingly, the condition on $g_{mM}$ decouples. Hence binding cannot be exclude if $g_{mM} > 1/2$. If $g_{mM} < 1/2$, then $g_{mm}, g_{MM}$ should lie outside the parabola shown if Fig. 10. To get a genuine Borromean state, one should exclude 3-body binding. If we replace the actual stability frontier for $(mmM)$ or $(MMm)$ by the bound (3.15), then we get a strict upper bound on the size of domain for a 4-body Borromean system. This is shown in the second part of Fig. 10. The floor of the volume is either the $g_{mM} = 1/2$ limit, or $g_{mM} = 0$ if $g_{mm}$ and $g_{MM}$ are large enough, i.e., outside the parabola. The ceiling is the lowest of the parabolic cylinders related to the stability of $(mmM)$ and $(MMm)$.

---

[2] A slight error in the drawing of Fig. 2 of that reference and in its caption are corrected below.

Finally, there are many related problems. Let us mention one of them, which was discovered in the course of our investigations, and while reading Ref. [36]. Consider for instance our curve in Fig. 11. When it crosses the line $g' = 1$, it continues in the $(g < 1, g' > 1)$ region, and becomes a separation between a region on the left where only the (1,2) pair is bound, and a region on the right where 3-body bound states also occur. We have seen that the line is convex inside the unit square (unlike the schematic drawing in Ref. [36]). Above this square, it tends to a vertical straight line $g = 3/8$, corresponding to particle 3 interacting with a point (1,2) object. (Remember the normalization: $g > 1$ is required to bind a reduced mass $1/2$ in a potential $v$. Then binding a reduced mass $2/3$ in a potential $2v$ requires $g > 3/8$.)

The way of approaching this $g = 3/8$ limit can be studied using a variational wave function of the type

$$\Psi = \psi_{1S}(r)\xi(R) \tag{5.10}$$

with $r = |\vec{r}_1 - \vec{r}_2|$, and $R$ is the separation between particle 3 and the c.o.m. of (1,2). The effective potential governing $\xi(R)$ results from a simple integral over the (1,2) distribution. The value $2v(R)$ in the limit where (1,2) is point-like, receives an attractive correction if $\Delta v < 0$ at large distances, where $\Delta$ is the Laplacian operator. This corresponds to the dotted line in the schematic drawing in Fig. 11. It means for some potentials, that there is perhaps an absolute minimum $g_{\min} < 3/8$ for the coupling constant $g$ to get 3-body binding, and that this minimum is reached at some finite value of $g'$. Drawing with precision all separation curves in the whole $(g, g')$ domain, and studying their curvature properties would be of interest. There are certainly many exciting effects to be unraveled in the transition from separate clusters to collective binding.

## VI. SUMMARY

In this paper, we have discussed the properties of the domain of coupling constants which bound a 3-body system but leave all 2-body subsystems (1,2), (2,3) and (3,1) unbound. A lower limit on these coupling constants is obtained from a variant of the Hall–Post inequalities, i.e., from a systematic decomposition of the (1,2,3) Hamiltonian in 2-body Hamiltonians. This lower limit appears a simple consequence of the variational principle, and is independant of the shape of the potential, once the coupling constants are properly normalized. In the case of identical bosons, the critical couplings $g_3$ for the 3-body binding and $g_2$ for 2-body binding are such that their ratio $g_3/g_2$ cannot be less than $2/3$.

An upper limit, and approximate estimate, of this domain can be computed with variational methods for each specific interaction. This is done here in the case of a Yukawa potential. There is typically a 20% window on the value of the coupling constants for binding the (1,2,3) system without 2-body binding. For instance, $g_3/g_2 \simeq 0.804$ for identical bosons.

Some of our rigourous results can be generalized to situations involving more than three particles, but the corresponding numerical estimates remain to be done. The case of identical fermions with attractive interaction would also require new investigations.

## ACKNOWLEDGMENTS

The Institut de Physique Nucléaire is supported by Université Claude Bernard, the Institut des Sciences Nucléaires by Université Joseph Fourier, and both by CNRS–IN2P3. J.M.R. would like to thank the hospitality of the Bonn University, and the support provided by an Alexander von Humboldt French–German Research Grant. We would like to thank B. Metsch for useful advice, and A.J. Cole for comments on the manuscript.

TABLES

TABLE I. Ground-state energy $E$ at given coupling $g$, and minimal coupling $g_3$ required to achieve binding, for a symmetric 3-body system of constituent masses $m_i = 1$ interacting through a Yukawa potential of range set to unity. The values are computed with different methods described in the text. For the Gaussian or exponential expansion, we use $G = 4$ terms.

| Method | $E(g = 1.7)$ | $E(g = 1.8)$ | $g_3$ |
|---|---|---|---|
| Hyperscalar | $-0.1141$ | $-0.2008$ | 1.491 |
| Feshbach–Rubinow | $-0.1841$ | $-0.2851$ | 1.390 |
| Gaussian expansion | $-0.1785$ | $-0.2780$ | 1.398 |
| Exponential expansion | $-0.1894$ | $-0.2897$ | 1.362 |
| Faddeev $l = 0$ | $-0.1893$ | $-0.2823$ | 1.365 |

# List of Figures

FIGURES

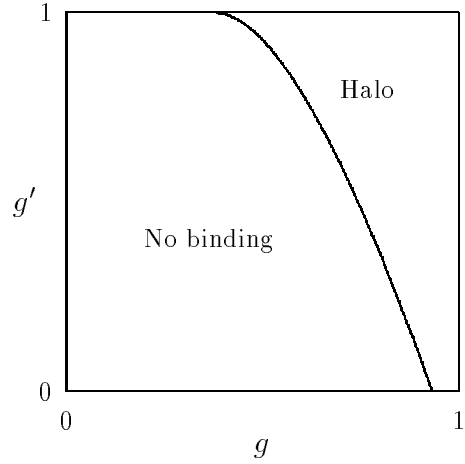

FIG. 1. Expected shape for the domain of Borromean binding inside the unit square of normalized coupling constants ($g \leq 1, g' \leq 1$).

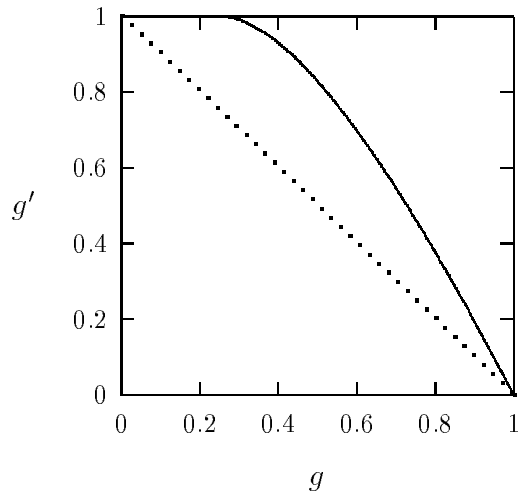

FIG. 2. Lower bound for the frontier of Borromean stability, for particules with masses $(1, 1, \infty)$. The coupling constants are normalized so that $g' = 1$ is required to bind particles 1 and 2 together and $g = 1$ to bind 1 or 2 around the static centre. The dotted line comes from the simple decomposition (3.10) of the Hamiltonian, the solid curve from Eq. (3.15).

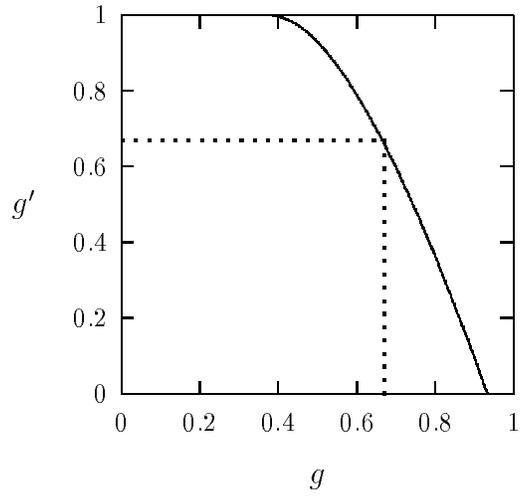

FIG. 3. Lower bound for the frontier of Borromean stability, for particles with unit masses, and coupling constant $g$ for the (1,3) and (2,3) interaction, and $g'$ for (1,2) one, normalized so that $g = 1$ or $g' = 1$ is the threshold for 2-body binding. The dotted square comes from the simple decomposition (3.11) of the Hamiltonian, the solid curve from Eq. (3.15).

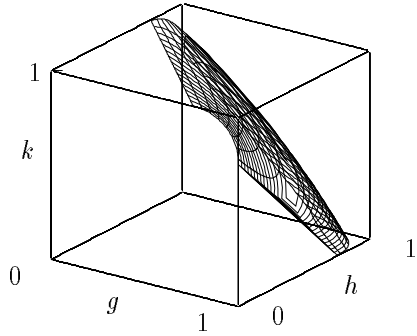

FIG. 4. Lower limit for the normalized coupling constants $g$, $h$ and $k$ to form a Borromean bound state. The figure corresponds to equal masses.

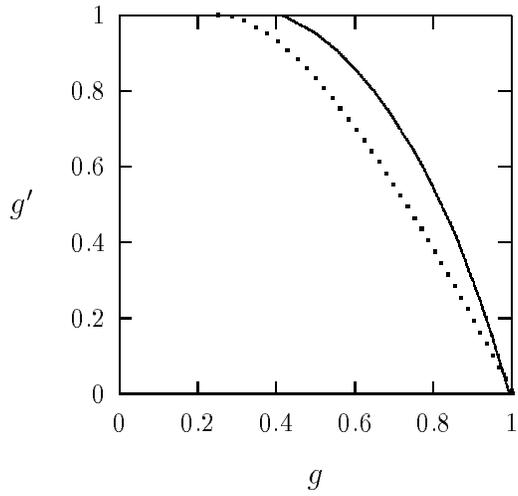

FIG. 5. Numerical estimate, in the space of normalized coupling constants $g$ and $g'$, of the frontier to form a Borromean bound state. The particles, of masses $m_1 = m_2 = 1$, $m_3 = \infty$, interact through a Yukawa potential. The dotted line is the rigourous lower limit.

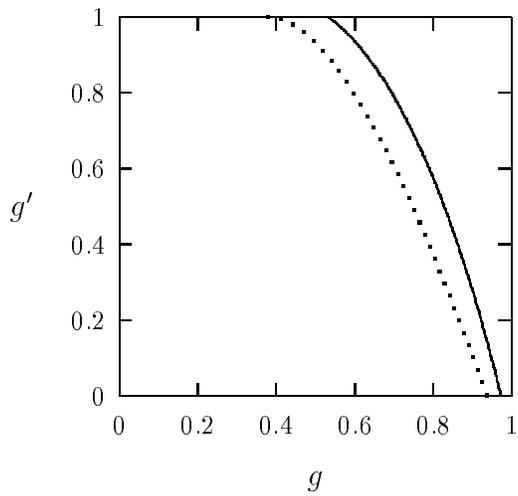

FIG. 6. Numerical estimate, in the space of normalized coupling constants $g$ and $g'$, of the frontier to form a Borromean bound state. The particles have equal masses and interact through a Yukawa potential. The dotted line is the rigourous lower limit.

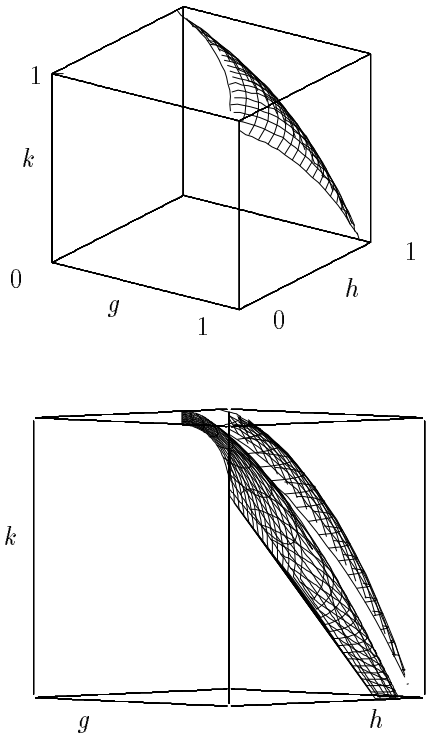

FIG. 7. Numerical estimate, in the space of normalized coupling constants $g$, $h$ and $k$, of the frontier to form a Borromean bound state. The plot shown here corresponds to equal masses interacting through a Yukawa interaction. The second figure is a tentative comparison of this frontier with the rigorous lower limit.

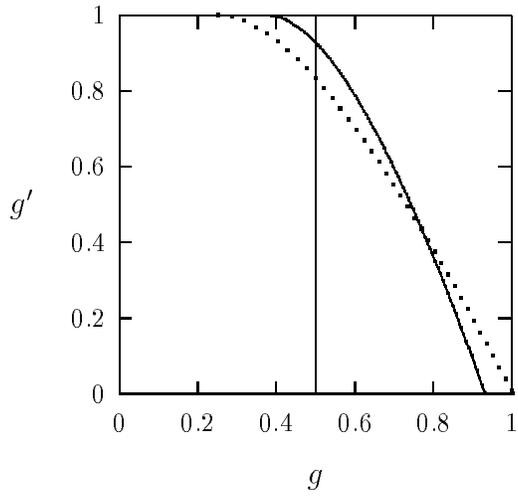

FIG. 8. Comparison of the maximal extension of the Borromean domain for a 3-body systems of masses $(m, m, M)$, in the cases $M \gg m$ (dotted line), $M = m$ (solid line), and $M \ll m$ (vertical straight line).

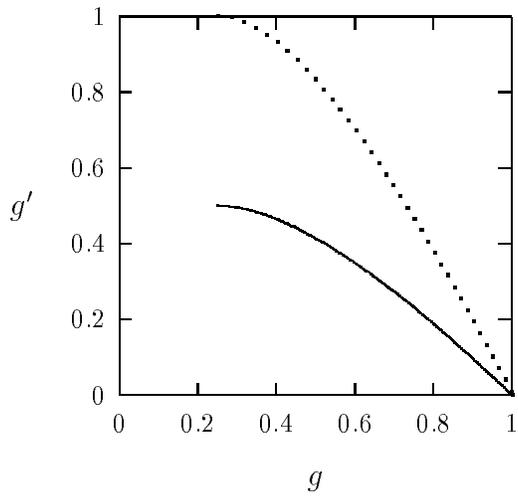

FIG. 9. Limit on the frontier of stability, for $p = 3$ bosons around a fixed centre, as deduced by Eq. (5.5) for the frontier of $p = 2$ bosons, shown as a dotted line.

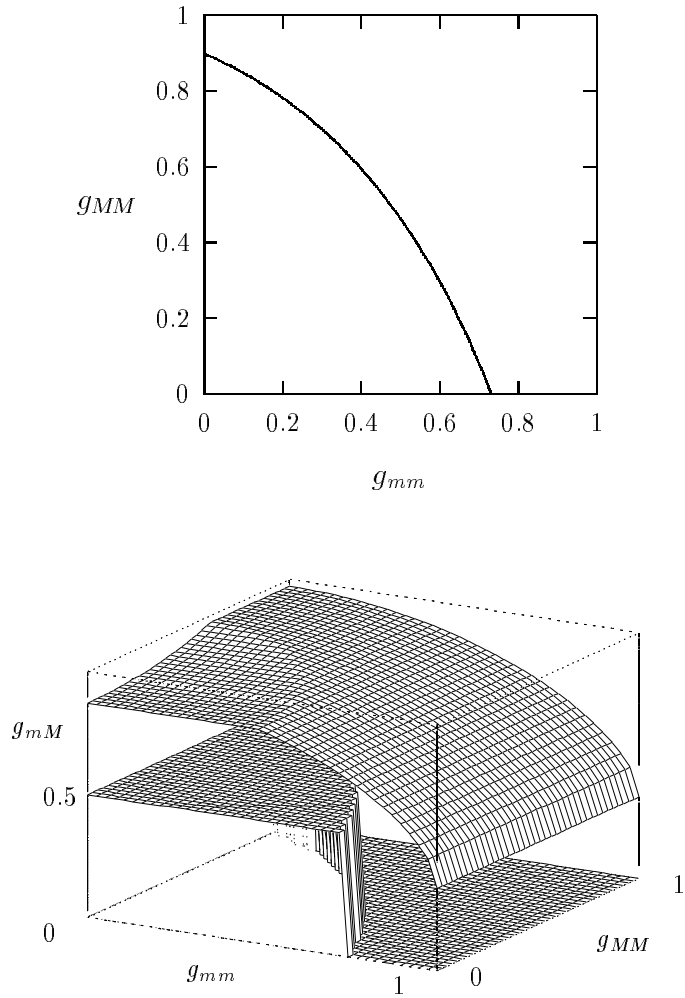

FIG. 10. Mixed limits on normalized coupling constants $g_{mm}$, $g_{MM}$ and $g_{mM}$ to form a 4-body Borromean state $(m, m, M, M)$. If $g_{mM} > 1/2$, binding cannot be excluded. If $g_{mM} < 1/2$, $g_{mm}$ and $g_{MM}$ should not correspond to a point inside the parabola. A value $M/m = 2$ is assumed here for the drawing. The 3-dimensional plot shows this lower limit on $g_{mM}$, and also a lower bound to the maximal value of $g_{mM}$: the intersecting cylinders are the lower bounds on $g_{mM}$ for binding $(m, m, M)$ or $(M, M, m)$.

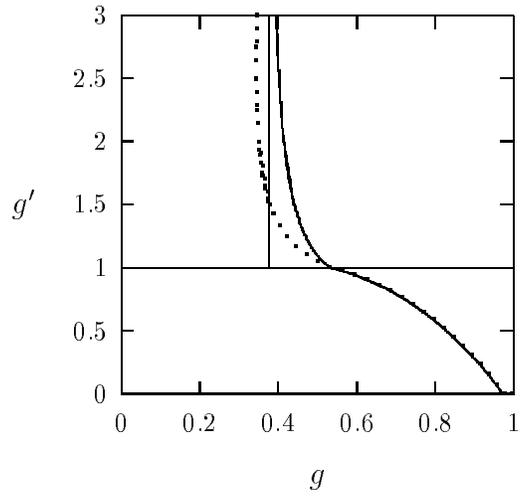

FIG. 11. Tentative extrapolation of the stability curve outside the unit square $g \leq 1, g' \leq 1$. The upper part separates the region where only (1,2) binding is permitted from the region where a 3-body bound state also exists. The vertical line corresponds to the limit where the (1,2) pair form a point-like cluster. The solid and dotted lines are two possible approaches to this limit.